\documentclass[journal,12pt,onecolumn,draftclsnofoot,]{IEEEtran} 
\ifCLASSINFOpdf
\else
\fi

\usepackage{graphicx}
\usepackage{amsmath}
\usepackage{amsthm}
\usepackage{amssymb}
\usepackage{algpseudocode}
\usepackage{cite}
\usepackage{color}
\usepackage{multirow}
\usepackage{mathptmx}
\usepackage{autobreak}
\newcommand\numberthis{\addtocounter{equation}{1}\tag{\theequation}}

\begin{document}

\title{\LARGE Virtual User Pairing Non-Orthogonal Multiple Access\\ in Downlink Coordinated Multipoint Transmission}
\author{Denny Kusuma Hendraningrat, Muhammad Basit Shahab, and Soo~Young~Shin,~\IEEEmembership{Senior Member,~IEEE} \vspace{-3.5ex}

	\thanks{Denny Kusuma Hendraningrat, Muhammad Basit Shahab, and Soo Young Shin are with the WENS Laboratory, Department of IT Convergence Engineering, Kumoh National Institute of Technology, Gumi 39177, South Korea
		(email: dennykh@ieee.org, basit.shahab@kumoh.ac.kr, and wdragon@kumoh.ac.kr).}%
}

\markboth{IEEE Communications Letters,~Vol.~xx, No.~xx, xx~2019}%
{}
\maketitle
\begin{abstract}
\noindent In this paper, joint transmission coordinated multipoint (JT-CoMP) is exploited by using virtual user pairing non-orthogonal multiple
access (VP-NOMA), termed as JT-CoMP VP-NOMA. The technique combines both VP-NOMA for enhancing ergodic sum capacity (ESC) and JT-CoMP for inter-cell interference mitigation. To show the performance gains, ESC of a three-cell scenario is analyzed as a key performance metric. The analytical and simulation results of JT-CoMP VP-NOMA are compared with orthogonal multiple access (OMA), non-orthogonal multiple access (NOMA), and VP-NOMA. It is shown that the proposed JT-CoMP VP-NOMA outperforms the other schemes in the viewpoint of ESC.
\end{abstract}
\IEEEpeerreviewmaketitle
\section{Introduction}
\IEEEPARstart{F}{IFTH} generation of cellular technology (5G) is the next phase of mobile telecommunication standard which aims at higher performance gains compared to existing 4G [1],[2]. Among many candidate technologies for 5G, non-orthogonal multiple access (NOMA) has recently gained much research interest as a candidate MA technique for 5G and beyond, which focuses on improving the spectrum efficiency by supporting multiple users over a particular channel resource, unlike conventional orthogonal MA (OMA) schemes [3]-[5]. The signals of multiple users are superimposed in the power domain, where successive interference cancellation (SIC) is used at the receivers for data recovery [6],[7]. Most of the works on NOMA pair a near user with a far user, where equal number of near and far users is assumed, so that each user can find a pair. However, the distribution of users in a cell is random in general; number of far users can be more than near users or vice versa. In [8]-[10], a virtual user pairing based NOMA (VP-NOMA) was suggested by considering a scenario with more far users than near users. In such scenario, VP-NOMA pairs a single near user with multiple far users over non-overlapping frequency bands. The analysis however was performed for a single cell scenario.
\par
In multi-cell scenarios, inter-cell interference is one of the critical issues as it degrades the system performance i.e., user throughput and cell capacity. Therefore, inter-cell interference mitigation techniques are critical in multi-cell scenarios [11]-[16]. In this context, joint transmission coordinated multi-point (JT-CoMP) has emerged as an important technology, where multiple BS in the neighboring cells cooperate with each other to support a common cell edge (far) user, thereby improving its achievable throughput. 
\par
The existing works on JT-CoMP in multi-cell scenarios consider one near user in each neighboring cell, while a single common far user. This is true if we assume less number of far users compared to the near users in the network, which is not always true. By considering a three-cell NOMA scenario, where each neighboring cell has a NOMA pair (a near and a far user), such that the system model now contains three near and three far users, a JT-CoMP VP-NOMA scheme is proposed to enhance the performance of JT-CoMP.
\par
The main contributions of this paper are listed as follows.
\begin{itemize}
\item This paper proposes a JT-CoMP VP-NOMA scheme by applying JT-CoMP on a generalized system model where each neighboring cell employs VP-NOMA. 
\item As conventional CoMP systems consider one common far user supported by neighboring BS [13],[14], the proposed work supports multiple users located in the cell edge, which is more suitable considering NOMA.
\item Closed-form solutions of ESC for the proposed JT-CoMP VP-NOMA are analyzed, and compared with OMA, NOMA, and VP-NOMA by considering imperfect SIC [17] and imperfect channel state information (CSI) [18].
\end{itemize}
The rest of this paper is organized as follows: Section II presents the considered system model for JT-CoMP VP-NOMA, while the associated transmission protocol is explained in Section III. Closed-form solution for the ESC through comprehensive mathematical analysis is provided in Sec IV. Section V provides simulation results and discussion. Finally, Section VI concludes the paper.
\section{System Model}
In [11]-[16], CoMP as an interference avoidance scheme based on cell coordination is discussed. In addition, JT-CoMP coordinates more than one transmit BSs to serve a specific user [13]-[16]. So, it is expected not only to mitigate inter-cell interference for CoMP-users but also may increase capacity performance.
\par
\begin{figure}[!t]
\centering
    \includegraphics[width=8.5cm,height=7.5cm]{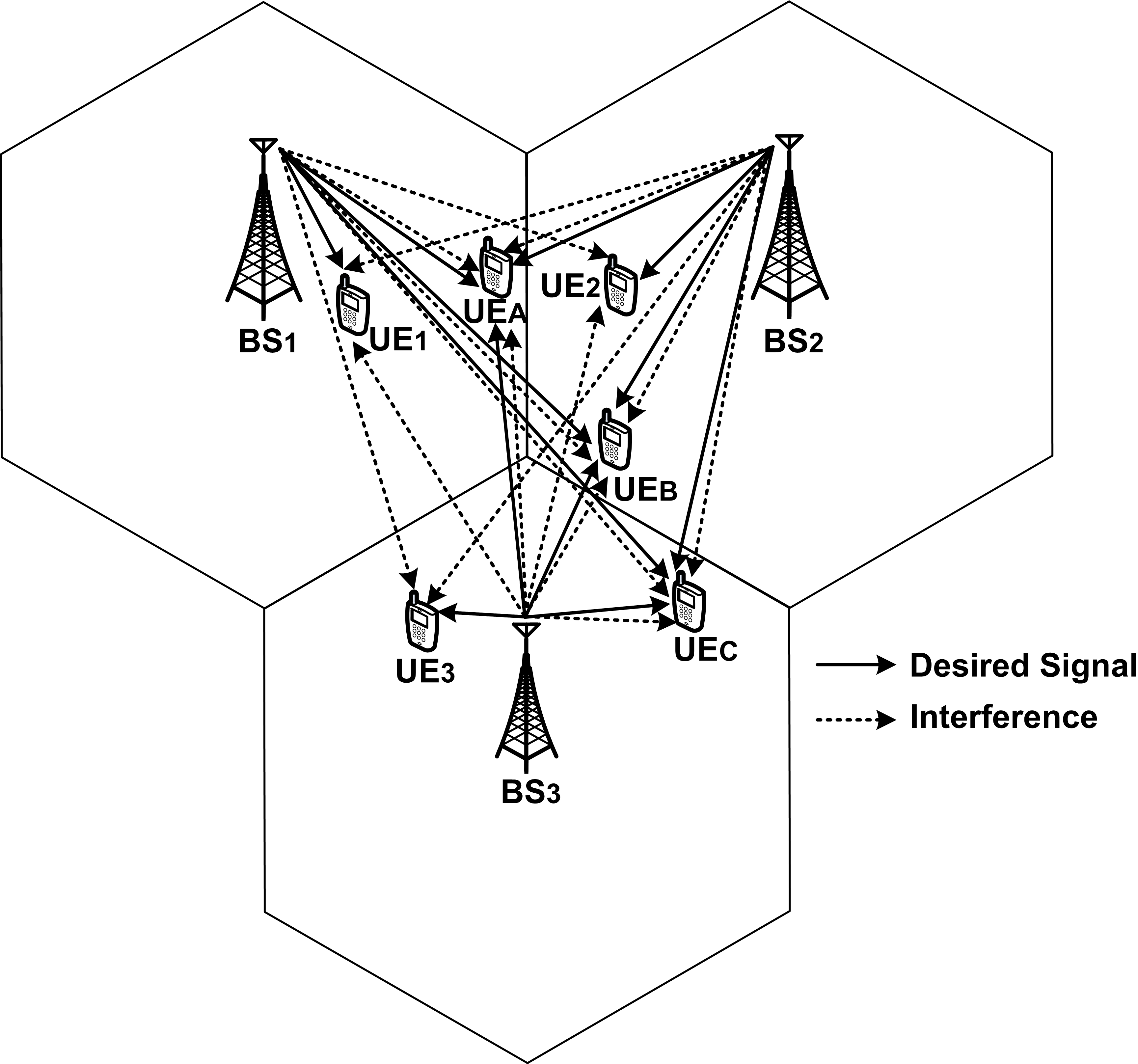}
\caption{System model}
\label{fig1}
\end{figure}

This paper proposes JT-CoMP VP-NOMA by combining VP-NOMA [8]-[10] and JT-CoMP for a NOMA cluster [16]. In the system model, we consider a three-cell scenario, where each cell consists of a BS, one near user and one far user; cell-1$\rightarrow$($BS_1$, $UE_1$, $UE_A$), cell-2$\rightarrow$($BS_2$, $UE_2$, $UE_B$), cell-3$\rightarrow$($BS_3$, $UE_3$, $UE_C$), as shown in Fig. \ref{fig1}. The users with integer subscripts are near users, whereas those with alphabetical subscripts are far users. The BS to user distance is defined as $r_{ij}$, where $i\in$1,2,3 represents the $i^{th}$ BS, and $j\in$1,2,3 represents the $j^{th}$ user [19]. Furthermore, $r_{ik}$ is $UE_k$ ($k\in$A,B,C) distances from $BS_i$. The maximum cell radius is normalized to $R=1$. 
The distances $d_{ij}$ and $d_{ik}$ between users and BS antenna are calculated by using the concepts of trigonometry [19].
\par
Based on [8]-[10], we implement a virtual user pairing scheme by pairing a near user $UE_{j}$ with three far users $UE_A$, $UE_B$, and $UE_C$.  In the system model, we consider that the three far users receive signals from all BSs. 
\section{Transmission Protocol}
\begin{figure}[!t]
	\centering
	\includegraphics[width=8.5cm,height=7.5cm]{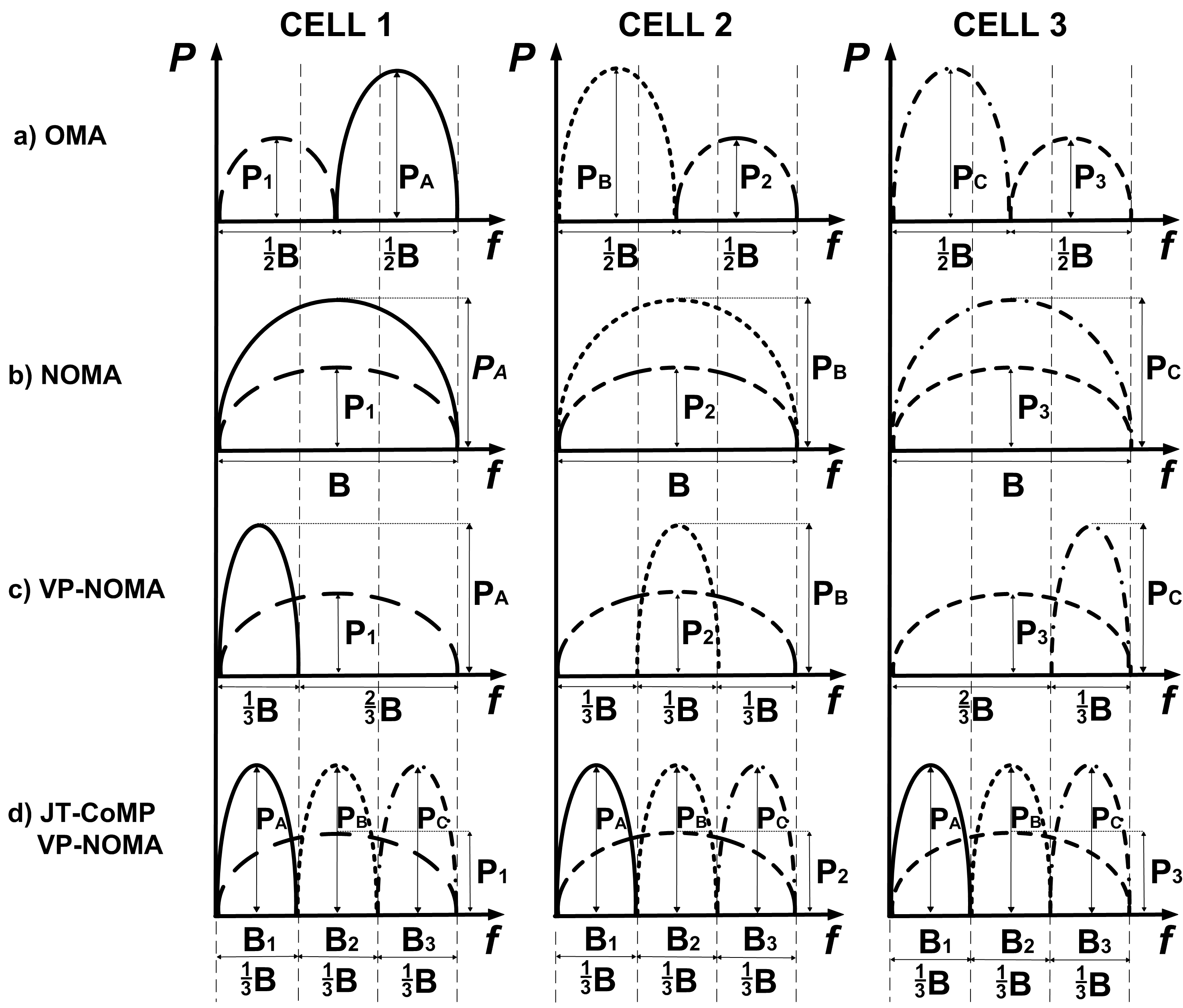}
	\caption{Considered scenarios: (a) OMA, (b) NOMA, (c) VP-NOMA, and     (d) JT-CoMP VP-NOMA}
	\label{fig2}
\end{figure}

\noindent Considering the system model in Fig. \ref{fig1}, transmission protocol of proposed scheme is shown in Fig. \ref{fig2}d. Each BS pairs its near user with all the three far users using VP-NOMA, such that near user gets the whole bandwidth while the three far users get non-overlapping parts of the whole bandwidth [8]-[10]. The same pattern is repeated in all the neighboring cells. In each cell, the near user performs SIC to recover its message signal, where as far users try to directly decode their signals by treating near user's low power signal as noise. Since perfect knowledge of CSI is not always possible, so imperfect CSI and correspondingly imperfect SIC process need to be considered at each $j^{th}$ near user. Imperfect CSI is modeled with channel estimation error, where a priory of variance of the error estimation is known. The channel estimation error for the near user can be modeled as $h_{\epsilon ij}$=$h_{ij}$-$\hat{h}_{ij}$. In addition, channel estimation error for the far user can be written as $h_{\epsilon ik}$=$h_{ik}$-$\hat{h}_{ik}$. It is assumed channel over each link is independent Rayleigh flat fading with channel coefficients $h_{\epsilon ij}\sim CN\left(0,\sigma_{\epsilon ij}\right)$, and $h_{\epsilon ik}\sim CN\left(0,\sigma_{\epsilon ik}\right)$.
\par
In wireless transmission, received power for the $UE_j$ need to consider as channel estimation gain $\left | \hat{h}_{ij} \right |^{2}$, where $\hat{h}_{ij}$ represents a channel estimation characteristic from the $BS_i$ antenna to the $UE_j$. In this paper, the channel estimation characteristic for the near user can be modeled as $\hat{h}_{ij}\sim CN\left(0,\hat{\sigma}_{ij}=d_{ij}^{-v}-\sigma_{\epsilon ij}\right)$ with mean zero and estimation variance $\hat{\sigma_{ij}}$ for the link from $BS_i$ antenna to $UE_j$, where $\mathit{v}$ represents the path-loss exponent and $\sigma_{\epsilon ij}$ represents the variance of the error estimation parameter for the link from $BS_i$ antenna to $UE_j$. In addition, we need consider $\left | \hat{h}_{ik} \right |^{2}$ as a channel estimation gain from $d_{ik}$, where channel estimation characteristic for the far user can be written as $\hat{h}_{ik}\sim CN\left(0,\hat{\sigma}_{ik}=d_{ik}^{-v}-\sigma_{\epsilon ik}\right)$. 
\par
Considering cell-1, if we normalize  total bandwidth with $B$=1, then $B_{1}=B_{2}=B_{3}=\frac{B}{3}$, such that $B_{1} \cap B_{2} \cap B_{3}=\O$, and $B_{1}+B_{2}+B_{3}=B=1$. Furthermore, we normalize total transmit power of BS by $P=1$, and denote transmit SNR by $\rho =\frac{P}{N_o}$, where $N_o$ represents AWGN. In this scenario, the BS power is divided for four users. If we assign $\alpha$ as a power allocation factor for the near user (non-CoMP user) in each cell, and $\beta$ as a power allocation factor for the far user (CoMP user) in each cell, then $\beta=\frac{(1-\alpha)}{3}$, where $\beta$$>$$\alpha$ and $\alpha+3\beta\leq1$. In addition, $P_j$=$\alpha$$P$ and $P_k$=$\beta$$P$ represent power allocation for the near and far user, respectively.
\section{Ergodic Sum Capacity Analysis}
\noindent In this section, we determine the ESC performance of JT-CoMP VP-NOMA scheme by considering imperfect CSI and imperfect SIC. First, we calculate the achievable data rate of the near user in the sub-carrier $B_1$ as follows:
\begin{align*}
C_{j\rightarrow B_1} &= B_{1}\log_2 \left( 1+ \frac{\alpha \rho |\hat{h}_{jj}|^2}{\alpha \rho \sum\limits_{\substack{i=1 \\ i \neq j}}^{3}|\hat{h}_{ij}|^2 + \rho \sum\limits_{i=1}^{3} \sigma_{\epsilon_{ij}}+\rho \Upsilon + 1 } \right)\\
&= B_{1}\log_2 \left( \frac{\alpha \rho \sum\limits_{i=1}^{3}|\hat{h}_{ij}|^2 + \rho \sum\limits_{i=1}^{3} \sigma_{\epsilon_{ij}}+\rho \Upsilon +1}{\alpha \rho \sum\limits_{\substack{i=1 \\ i \neq j}}^{3}|\hat{h}_{ij}|^2 + \rho \sum\limits_{i=1}^{3} \sigma_{\epsilon_{ij}} + \rho \Upsilon + 1 }\right)\numberthis \label{eqn}
\end{align*}
where $\gamma$ denotes residual interference, representing imperfect SIC at the near user.
\par
By using $\log_n(x/y) = \log_n(x)-\log_n(y)$, (1) can be rewritten as
\begin{align*}
C_{j\rightarrow B_1}
&= B_{1}\left\{\log_2 \left( \alpha \rho \sum\limits_{i=1}^{3}|\hat{h}_{ij}|^2 + \rho \sum\limits_{i=1}^{3} \sigma_{\epsilon_{ij}} + \rho \Upsilon +1 \right) \right.\\
&\left.\ \ \ -
\log_2 \left(  \alpha \rho \sum\limits_{\substack{i=1 \\ i \neq j}}^{3}|\hat{h}_{ij}|^2 + \rho \sum\limits_{i=1}^{3} \sigma_{\epsilon_{ij}} + \rho \Upsilon + 1 \right)\right\}.\numberthis \label{eqn}
\end{align*}
In addition, the achievable data rate of $UE_{A}$, getting signals from all three BSs over the bandwidth $B_1$, can be calculated as
\begin{align*}
C_{A\rightarrow B_1}&=B_{1}\log_{2}\left(1+\frac{\beta\rho\sum\limits_{\substack{i=1}}^{3}\left | \hat{h}_{iA} \right |^{2}}{\alpha\rho\sum\limits_{\substack{i=1}}^{3}\left | \hat{h}_{iA} \right |^{2}+\rho\sum\limits_{\substack{i=1}}^{3}\sigma_{\epsilon{iA}}+1}\right).\numberthis \label{eqn}
\end{align*}
Hence, the achievable sum rate $C_{123A}^{erg}$ of sub-carrier $B_1$  is given as
\begin{equation}
C_{123A}^{erg}=\sum_{j=1}^{3}C_{j\rightarrow B_1}+C_{A\rightarrow B_1}.
\end{equation}
Similarly, we can determine achievable sum rate in sub-carrier $B_2$ and $B_3$. Therefore, total ESC for all users is given as
\begin{equation}
C_{total}^{erg}=C_{123A}^{erg}+C_{123B}^{erg}+C_{123C}^{erg}    
\end{equation}
Furthermore, from (2), the exact ergodic capacity $C_{j\rightarrow B_1}^{exact}$ in sub-carrier $B_1$ can be calculated by solving
\begin{align*}
C_{j\rightarrow B_1}^{\text{exact}} &= E\{C_{j\rightarrow B_1}\} \\
&= B_{1}\left\{ \int_{0}^{\infty} \log_2 \left( x + a \right) f_{X_j}(x)dx-\right. \\
&\ \ \ \left. \int_{0}^{\infty} \log_2 \left( y + a \right) f_{Y_j}(y) dy \right\},\numberthis \label{eqn}
\end{align*}
where $E$ is expectation operator, and $a=\rho \sum_{i=1}^{3} \sigma_{\epsilon_{ij}}+\rho\gamma+1$.
\par
Therefore, by using the probability density function of $f_{X_j}(x)$ and $f_{Y_j}(y)$ which are derived in [19], $C_{j\rightarrow B_1}^{\text{exact}}$ in sub-carrier $B_1$ can be calculated as
\begin{align*}
C_{j\rightarrow B_1}^{\text{exact}}
&= \frac{B_{1}}{\ln(2)} \left\{ \sum\limits _{i=1}^{3} \left( \ln(a) - \exp(ak_{ij}) \textrm{Ei}(-ak_{ij}) \right) \times \prod\limits_{\substack{h=1 \\ h \neq i}}^{3} \frac{k_{hj}}{k_{hj} - k_{ij}} \right.\\
&\left. \ \ - \sum\limits _{\substack {i=1 \\ i \neq j}}^{3} \left( \ln(a) - \exp(ak_{ij}) \textrm{Ei}(-ak_{ij}) \right) \times \prod\limits_{\substack{h=1 \\ h \neq i \\ h \neq j}}^{3} \frac{k_{hj}}{k_{hj} - k_{ij}} \right\}, \numberthis \label{eqn}
\end{align*}
where $k_{ij} = \frac{1}{\alpha \rho \hat{\sigma}_{ij}}$ and $k_{hj} = \frac{1}{\alpha \rho \hat{\sigma}_{hj}}$.
\par
Similarly, by assuming $b=\rho \sum_{i=1}^{3} \sigma_{\epsilon_{ij}}+1$, we can calculate the achievable data rate for $UE_{A}$ as
\begin{align*}
C_{A\rightarrow B_1}^{\text{exact}}
&= \frac{B_{1}}{\ln(2)} \left\{ \sum\limits _{i=1}^{3} \left( \ln(b) - \exp(bl_{iA}) \textrm{Ei}(-bl_{iA}) \right) \times \prod\limits_{\substack{h=1 \\ h \neq i}}^{3} \frac{l_{hA}}{l_{hA} - l_{iA}} \right.\\
&\left. \ \ - \sum\limits _{\substack {i=1}}^{3} \left( \ln(b) - \exp(bm_{iA}) \textrm{Ei}(-bm_{iA}) \right) \times \prod\limits_{\substack{h=1 \\ h \neq i}}^{3} \frac{m_{hA}}{m_{hA} - m_{iA}} \right\}, \numberthis \label{eqn}
\end{align*}
where $l_{iA} = \frac{1}{(\alpha + \beta) \hat{\sigma}_{iA}}$, $l_{hA} = \frac{1}{(\alpha + \beta) \hat{\sigma}_{hA}}$,  $m_{iA} = \frac{1}{\alpha\rho \hat{\sigma}_{iA}}$, and $m_{hA} = \frac{1}{\alpha \rho\hat{\sigma}_{hA}}$.
\par
By combining (4), (5), (7) and (8), ESC for all users can be calculated as
\begin{equation}
C_{total}^{exact}=C_{123A}^{exact}+C_{123B}^{exact}+C_{123C}^{exact}.    
\end{equation}.
\section{Simulation Results and Discussion}
\begin{figure}[!t]
\centering
    \includegraphics[width=7.75cm,height=6.5cm]{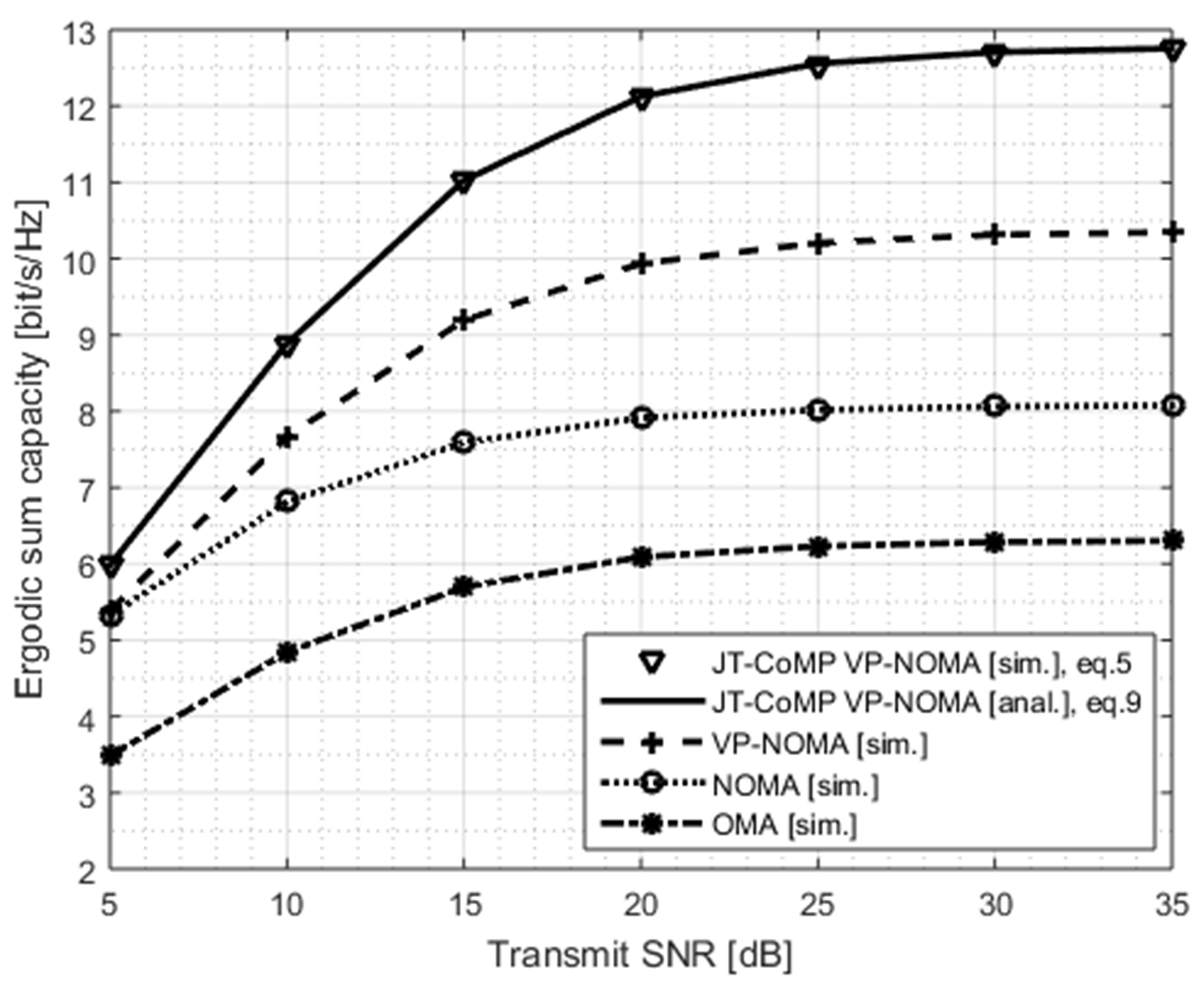}
\caption{Ergodic sum capacity with respect to transmit SNR ($\rho$), $\alpha$=0.1 and $r_{11}$=0.5R}
\label{fig3}
\end{figure}
In this section, we analyze the ESC of the proposed scheme in comparison with the conventional benchmark techniques. 
\par
Fig. 3 shows the simulation and analytical results for the ESC of JT-CoMP VP-NOMA in comparison with OMA, NOMA, and VP-NOMA. The results show that ESC of JT-CoMP VP-NOMA is higher than other schemes. This is because using both JT-CoMP and VP-NOMA together ensures better spectrum usage, and better interference management simultaneously, thereby enhancing the overall performance of the proposed scheme.
\begin{figure}[!t]
\centering
    \includegraphics[width=7.75cm,height=6.5cm]{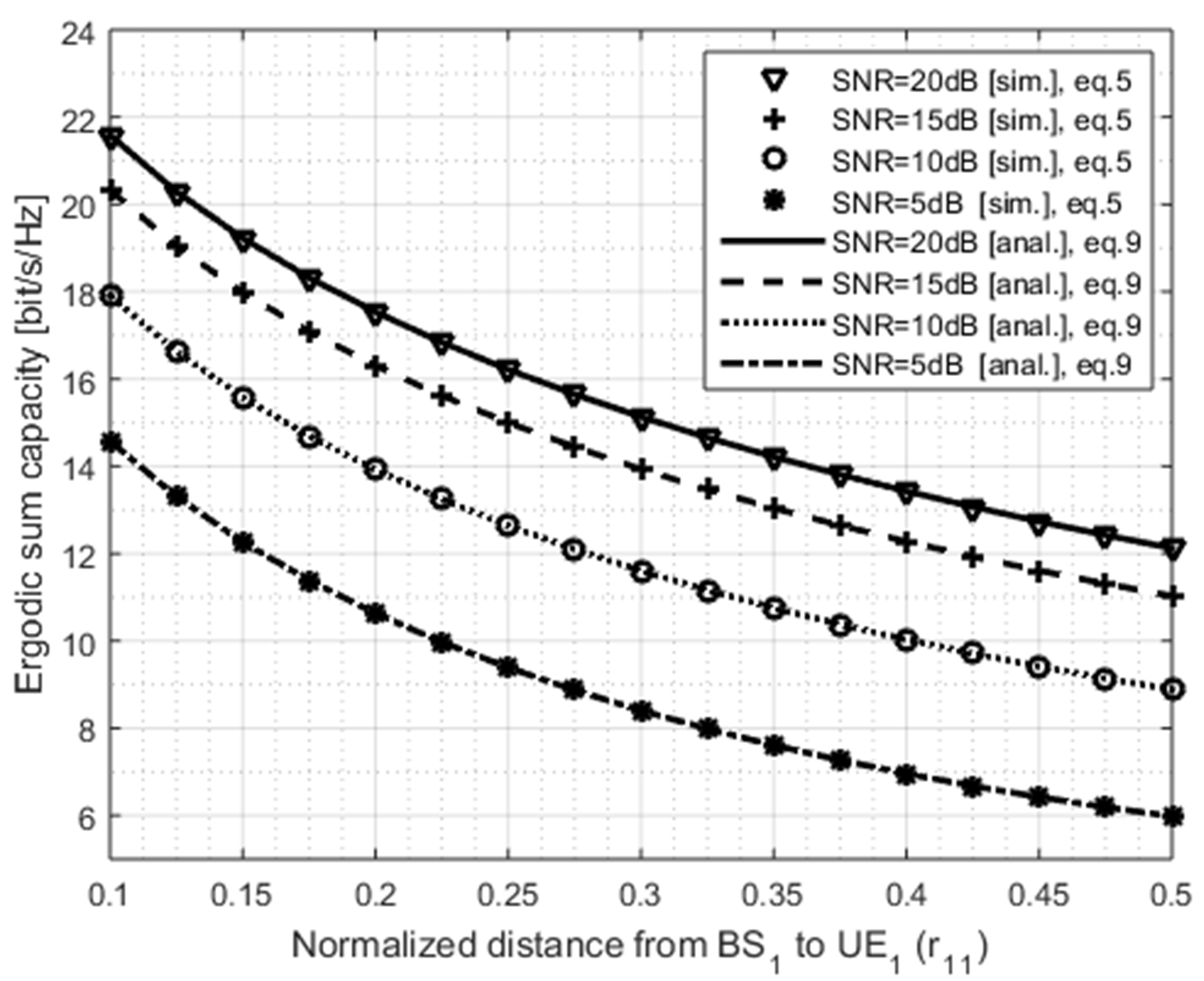}
\caption{Ergodic sum capacity of JT-CoMP VP-NOMA with respect to $UE_1$ distance from $BS_1$ ($r_{11}$), and $\alpha$=0.1}
\label{fig4}
\end{figure}
\begin{figure}[!t]
\centering
    \includegraphics[width=7.75cm,height=6.5cm]{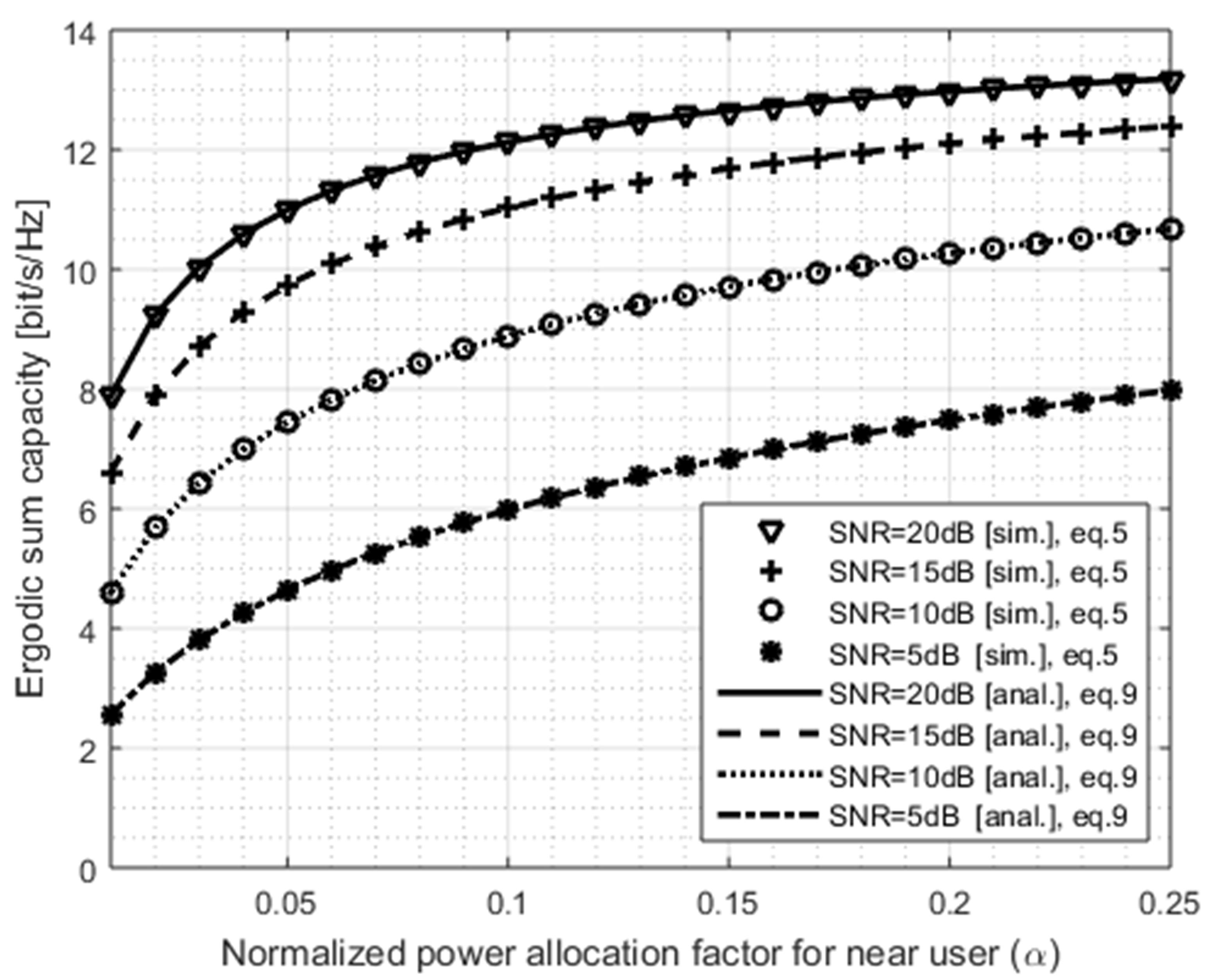}
\caption{Ergodic sum capacity of JT-CoMP VP-NOMA with respect to power allocated for the near user ($\alpha$), and $r_{11}$=0.5R}
\label{fig5}
\end{figure}
\par
Fig. 4 shows the effect of user locations on the performance of the proposed scheme. Far users are kept fixed. Near users are initially close to the BS, and then moved away from the BS (towards the far users). It can be seen that JT-CoMP VP-NOMA achieves higher ESC if the distance between near and far user is large. When the near user moves towards the cell edge, the system capacity gets degraded. This is due to increased mutual interference between users and channel degradation of near user. Finally, Fig. 5 shows the effect of power allocation factors on the ESC of the proposed scheme. The power allocation factor of near users is increased, which increases the overall capacity of the system, which is in line with the existing literature.
\section{Conclusion}
In this paper, we propose a JT-CoMP VP-NOMA scheme to increase the ESC of a multicell NOMA system. The simulation results show that proposed JT-CoMP VP-NOMA achieves higher ESC compared to OMA, NOMA, and VP-NOMA. Moreover, ESC for proposed system increases if the power allocated for the near user and distance between the near user and the far user are increased. 

\section*{References}

\noindent [1]   J. S. Park and B. J. Kim, “Trends and technical requirements for 5G mobile communication systems,”The Journal of the Korea institute of electronic communication sciences, vol. 10, no. 11, pp. 1257–1264, 2015.\\

\noindent [2]   K.  Tateishi  and  et.al,  “5G  experimental  trial  achieving  over  20  Gbps  using  advanced  multi-antenna  solutions,”  in 2016 IEEE 84th Vehicular Technology Conference (VTC-Fall), Sept 2016, pp. 1–5.\\

\noindent [3]   Y. Yang, J. Xu, G. Shi, and C. Wang,5G Wireless Systems-Simulation and Evaluation Techniques. Springer International Publishing, Switzerland, 2018.\\

\noindent [4]   Z.  Ding,  X.  Lei,  G.  K.  Karagiannidis,  R.  Schober,  J.  Yuan,  and  V.  K.  Bhargava, “A  survey  on  non-orthogonal  multiple access  for  5G  networks:  Research  challenges and  future  trends,”IEEE Journal on Selected Areas in Communications, vol. 35, no. 10, pp. 2181–2195, Oct 2017.\\

\noindent [5]   A. Benjebbour, A. L. K. Saito, Y. Kishiyama, and T. Nakamura,Signal Processing for 5G: Algorithms and Implementations,First Edition. John Wiley and Sons, Ltd, 2016.\\

\noindent [6]   M. B. Shahab, M. F. Kader, and S. Y. Shin, “On the power allocation of non-orthogonal multiple access for 5G wireless networks,” in 2016 International Conference on Open Source Systems Technologies (ICOSST), Dec 2016, pp. 89–94.\\

\noindent [7]   Y. Saito, Y. Kishiyama, A. Benjebbour, T. Nakamura, A. Li, and K. Higuchi, “Non-orthogonal multiple access (NOMA)for cellular future radio access,” in2013 IEEE 77th Vehicular Technology Conference (VTC Spring), June 2013, pp. 1–5.\\

\noindent [8]   M. B. Shahab, M. F. Kader, and S. Y. Shin, “A virtual user pairing scheme to optimally utilize the spectrum of unpaired users in non-orthogonal multiple access,” vol. 23, no. 12, Dec 2016, pp. 1766–1770.\\

\noindent [9]   M. B. Shahab and S. Y. Shin, “On the performance of a virtual user pairing scheme to efficiently utilize the spectrum of unpaired users in NOMA,”Physical Communication, vol. 25, no. 2, pp. 492–501, 2017.\\

\noindent [10]   M. F. Kader, M. B. Shahab, and S. Y. Shin, “Non-orthogonal multiple access for a full-duplex cooperative network with virtually paired users,”Computer Communications, vol. 120, no. 2, pp. 1–9, May 2018.\\

\noindent [11]   A. S. Hamza, S. S. Khalifa, H. S. Hamza, and K. Elsayed, “A survey on inter-cell interference coordination techniques in OFDMA-based cellular networks,” vol. 15, no. 4, Fourth 2013, pp. 1642–1670. IEEE COMMUNICATIONS LETTERS, VOL. XX, NO. XX, XX 201910\\

\noindent [12]   N.  Katiran  and  et.al,  “Inter-cell  interference  mitigation  and  coordination in  CoMP  systems,”  in Informatics Engineering and Information Science. ICIEIS 2011. Communications in Computer and Information Science, vol. 253, 2011.\\

\noindent [13]   J.  Choi,  “Non-orthogonal  multiple  access  in  downlink  coordinated  two-point  systems,”IEEE Communications Letters,vol. 18, no. 2, pp. 313–316, February 2014.\\

\noindent [14]   Y. Sun, Z. Ding, X. Dai, and G. K. Karagiannidis, “A novel network NOMA scheme for downlink coordinated three-point systems,”arXiv:1708.06498 [cs.IT], [Online]. https://arxiv.org/pdf/1708.06498.pdf, Dec 2017. \\

\noindent [15]   M. S. Ali, E. Hossain, A. Al-Dweik, and D. I. Kim, “Downlink power allocation for CoMP-NOMA in multi-cell networks,”IEEE Transactions on Communications, vol. 66, no. 9, pp. 3982–3998, Sept 2018.\\

\noindent [16]   M. S. Ali, E. Hossain, and D. I. Kim, “Coordinated multipoint transmission in downlink multi-cell NOMA systems: Models and spectral efficiency performance,”IEEE Wireless Communications, vol. 25, no. 2, pp. 24–31, April 2018.\\

\noindent [17]   M. B. Shahab, M. Irfan, M. F. Kader, and S. Y. Shin, “User pairing schemes for capacity maximization in non-orthogonal multiple access systems,”Wireless Communications And Mobile Computing, no. 16, pp. 2884–2894, Sept 2016.\\

\noindent [18]   H.  V.  Cheng,  E.  Bjrnson,  and  E.  G.  Larsson,  “NOMA  in  multiuser  MIMO  systems  with  imperfect  CSI,”  in2017 IEEE 18th International Workshop on Signal Processing Advances in Wireless Communications (SPAWC), July 2017, pp. 1–5.\\

\noindent [19]   F. W. Murti, R. F. Siregar, and S. Y. Shin, “Exploiting non-orthogonal multiple access in downlink coordinated multipoint transmission  with  the  presence  of  imperfect  channel  state  information,”Computer Sciences,  December  2018.  [Online].Available: https://arxiv.org/abs/1812.10266

\ifCLASSOPTIONcaptionsoff
  \newpage
\fi
\appendices

\bibliographystyle{IEEEtran}
\bibliography{IEEEabrv,}

%
\end{document}